\documentclass[showpacs,aps,amsmath,amssymb,twocolumn]{revtex4-2}

\usepackage[]{graphicx}
\usepackage{xcolor}
\usepackage{bbold}
\usepackage{braket}
\usepackage[colorlinks=true, pdfstartview=FitV, linkcolor=red, citecolor=blue, urlcolor=blue]{hyperref}
\newcommand{\be}{\begin{equation}}
\newcommand{\ee}{\end{equation}}
\newcommand{\ben}{\begin{eqnarray}}
\newcommand{\een}{\end{eqnarray}}
\newcommand{\bes}{\begin{subequations}}
\newcommand{\ees}{\end{subequations}}
\def\bal#1\eal{\begin{align}#1\end{align}}

\newcommand{\bfi}{\begin{figure}}
\newcommand{\efi}{\end{figure}}
\newcommand{\bc}{\begin{center}}
\newcommand{\ec}{\end{center}}

\newcommand{\sgn}{\text{sgn}}

\DeclareMathOperator{\arctanh}{arctanh}

\begin{document}

\title{{Cuscuton-like contribution to dark energy evolution}}
\author{D. Bazeia$^{1}$, J. D. Dantas$^{2}$, and S. Santos da Costa$^{3,4,5}$}
\affiliation{$^1$Departamento de F\'\i sica, Universidade Federal da Para\'\i ba, 58051-970 Jo\~ao Pessoa, Para\'\i ba, Brazil\\
$^2$Unidade Acadêmica de F\'{\i}sica e Matem\'atica, Universidade Federal de Campina Grande, 58175-000 Cuit\'e, Para\'{\i}ba, Brazil\\
$^3$Istituto Nazionale di Fisica Nucleare (INFN), Sezione di Pisa, Largo B. Pontecorvo 3, 56127, Pisa, Italy\\
$^4$Dipartimento di Fisica, Università di Trento, Via Sommarive 14, 38123 Povo, Trento, Italy\\
$^5$Trento Institute for Fundamental Physics and Applications (TIFPA), Via Sommarive 14, 38123 Povo, Trento, Italy
}

\begin{abstract} 
This work deals with the presence of the cuscuton term in the otherwise standard dark energy evolution under the usual FLRW background. We disclose a first-order framework similar to the Hamilton-Jacobi formalism, which helps us to solve the equations of motion and find analytical solutions. We explore several possibilities, concentrating mainly on how the cuscuton-like contribution works to modify cosmic evolution. Some results are of current interest since they describe scenarios capable of changing the evolution, adding or excluding possible distinct phases during the Universe's expansion history.  Additionally, we present interesting constraints on the cuscuton-like contribution for the dark energy evolution using a set of homogeneous geometrical observational probes. Finally, based on the Akaike Information Criterion (AIC), we perform a statistical comparison of the cuscuton-like model with $\Lambda$CDM, and find strong support for our model.
\end{abstract}

\maketitle

\section{Introduction}\label{introd}

Several cosmological observations point out that the universe is currently undergoing an accelerated expansion \cite{riess1998, perlmutter1999, knop2003, spergel2003, caldwell2004, riess2004, koivisto2006, daniel2008, kowalski2008, komatsu2009, eisenstein2005, percival2010}. Understanding the nature of the late-time cosmic acceleration is therefore among the main unsolved problems in Cosmology. In General Relativity (GR) this phenomenon is connected to the presence of an exotic energetic component with negative pressure called Dark Energy (DE), which dominates the energy density of the universe \cite{copeland2006, cai2010}. But this also leaves open the possibility of modified gravity. Horndeski \cite{horndeski1974,  deffayet2011, kobayashi2011} and Galileon \cite{zumalacarregui2013, gao2011, gleyzes2015a, gleyzes2015b, nicolis2009} models are examples of these alternative models. Modifications of Einstein's gravity, however, often give rise to extra dynamic degrees of freedom which should be also taken into account. 

The recent proposal of cuscuton gravity, on the other hand, appears as an infrared modification of GR without any additional degrees of freedom, as analyzed in \cite{gomes2017}. Cuscuton was first introduced in \cite{afshordi2007a, afshordi2007b} as a new DE model. More recently, the authors in \cite{iyonaga2018,iyonaga2020} have formulated extended cuscuton theories, also including extended cuscuton as dark energy. In addition to this, cuscuton models may provide interesting new features in several other contexts, such as Horava-Lifshitz gravity \cite{afshordi2009, bhattacharyya2018}, the ability of the cuscuton term to represent models of bouncing cosmology\cite{Boruah:2018pvq, Quintin:2019orx, Kim:2020iwq}, and to reconcile the power-law inflation model with CMB data~\cite{Ito:2019fie}. Moreover, multi-field cuscuton cosmology is also investigated in \cite{mansoori2022}, in \cite{CR} the investigation concerns the cuscuta–Galileon, which appears as a Galilean generalization of cuscuton field, in \cite{CG} the authors discuss cosmology based on a Cuscuta-Galileon gravity theory and in \cite{BW} the study deals with self-tuning of the cosmological constant within specific braneworld models.

In general, the equation of motion of the cuscuton field does not induce dynamics, so it just acts as an auxiliary field. Other recent possibilities appeared in \cite{bazeia2022a, bazeia2022b}, in the context of braneworld in the scalar-tensor representation of modified gravity in the presence of the cuscuton contribution. In these references, the authors extended the cuscuton Lagrangian by including the standard kinetic term, which can also be interpreted as adding the term cuscuton to the standard Lagrangian. 
This same type of extension has been carried out in \cite{zhong2021}, where the impacts of the cuscuton term in a dilaton gravity model are discussed.
In this context, we propose the introduction of a cuscuton-like term, to be added together with the standard kinematic contribution to the standard Lagrangian in the Einstein-frame, in a way similar to \cite{bazeia2022a, bazeia2022b, zhong2021}, to explore how much it contributes to the dynamics of dark-energy evolution.
 
To start this investigation pedagogically, in the next Sec. \ref{basic} we briefly review the basic formalism of Cosmology in the presence of a single real scalar field with the addition of a cuscuton-like term. The revision includes the equations of motion, energy density, pressure, and the equation of state. We go on and in Sec. \ref{formalism} we develop the first order framework in the presence of a single scalar field, and we illustrate the main results with two distinct models, with exponential and hyperbolic potentials. We add another scalar field in Sec. \ref{formalism_two} and generalize the first-order framework to this newer case, illustrating the results with four other models. In Sec. \ref{observa} we use observational probes to put constraints on the cuscuton contribution for the dark energy evolution.  We end the work in Sec. \ref{final}, summarizing the main results and suggesting new possibilities to explore the influence of the cuscuton-like contribution to the dark energy evolution.

\section{Basic framework}\label{basic}

Let us start considering the action
\be\label{action}
S=\int \mathrm{d}^4x \sqrt{-g} \left[\frac{1}{4} R +  \mathcal{L}(\phi,\partial_{\mu}\phi) \right],
\ee
where $R$ is the Ricci scalar, $\phi$ is a real scalar field, and we have adopted the standard values $4\pi G = c =1$. Varying the action with respect to $\phi$ yields the background equation of motion
\be\label{eq_motion}
\dfrac{1}{\sqrt{-g}}\partial_{\mu}\left(\sqrt{-g}\; \dfrac{\partial \mathcal{L}}{\partial (\partial_{\mu}\phi)}\right) - \dfrac{\partial \mathcal{L}}{\partial \phi} = 0.
\ee
For a flat, homogeneous, and isotropic universe, we take the flat FLRW line element
\begin{equation}\label{met_flrw}
    ds^2 = -dt^2 + a^2(t)\left(dr^2 + r^2d\Omega^2\right),
\end{equation}

\noindent where the expansion is described by the scale factor $a(t)$, $r$ is the radial coordinate and $d\Omega^2 = d\theta^2 + \sin^2\!\theta \;d\phi^2$ describes the angular portion of the metric.

We consider the extended cuscuton Lagrangian in the form
\be\label{lagrangian}
\mathcal{L} = - \frac{1}{2}\partial_{\mu}\phi\partial^{\mu}\phi - \alpha\sqrt{|\partial_{\mu}\phi\partial^{\mu}\phi|} - V(\phi),
\ee
where $\alpha$ is a positive real constant that accounts for the presence of the cuscuton-like term in the scalar field model. For the time-dependent $\phi=\phi(t)$, the equation of motion \eqref{eq_motion} becomes 
\be\label{eq_motion_2}
\ddot{\phi} + 3H(\dot{\phi} + \alpha\; \sgn(\dot{\phi})) + V_{\phi} = 0,
\ee

\noindent where $V_{\phi} = dV/d\phi$. The energy density and the pressure of the field are, respectively,
\begin{subequations}\label{rhoep}
\begin{equation}\label{rho}
    \rho = \frac{1}{2}\dot{\phi}^2 + V(\phi),
\end{equation}
\begin{equation}\label{p}
    p = \frac{1}{2}\dot{\phi}^2 + \alpha\; \dot{\phi}\; \sgn(\dot{\phi}) - V(\phi).
\end{equation}
\end{subequations}

\noindent We observe that the new cuscuton contribution does not change the energy density of the scalar field, which is the same as the standard dynamic. The pressure, however, is different from the standard case. In this sense, the equation of state (e.o.s) parameter, $\omega=p/\rho$, also changes, being given by
\be\label{eq_state}
\omega = \dfrac{\dot{\phi}^2 + 2\alpha\; \dot{\phi}\; \sgn(\dot{\phi}) - 2V}{\dot{\phi}^2 + 2V}.
\ee

\noindent From now on we will assume that $\sgn(\dot{\phi})>0$, and study the effect of the cuscuton-like term on cosmic evolution. In this scenario, the Einstein equations lead to the Friedmann equations
\begin{subequations}\label{frid_a}
\begin{equation}\label{hd_a}
    3H^2 \;=\; 2\rho \;=\; \dot{\phi}^2 + 2V,
\end{equation}
\begin{equation}\label{hp_a}
    \dot{H} \;=\; -(\rho + p) \;=\; - (\dot{\phi} + \alpha )\dot{\phi},
\end{equation}
\end{subequations}

\noindent and $\rho$ and $p$ satisfy the continuity equation
\be\label{continuity}
    \dot{\rho} + 3H\rho\left[1+\omega(t)\right] = 0.
\ee

The equation of state $\omega$ and the acceleration parameter $q$ are important parameters in the analysis of cosmological models and can be conveniently expressed in terms of $H$ and $\dot{H}$, as follows
\begin{equation}\label{omega_q_z}
    \dfrac{3}{2}(1 + \omega) \;=\; 1 - q \;=\; - \frac{\dot{H}}{H^2}.
\end{equation}
The expressions \eqref{omega_q_z} allow us to analyze different parametrizations in search of agreement with the current observational data. The condition $-1 \leq \omega < -1/3$ is required to realize the late-time cosmic acceleration. This imposes on $q$ the limits $0 < q \leq 1$. To study the parameters $\omega$ and $q$, we need to solve the second order differential equations (\ref{eq_motion_2}) and (\ref{hp_a}). So, in the next sections, we will develop procedures to help us obtain solutions in several distinct scenarios.

Before presenting the first-order formalism, evaluating whether or not this model respects the stability conditions is important. This analysis can be performed by studying the linear perturbations around an FLRW background and finding the conditions under which the propagation speeds of the tensor and scalar modes are positive, i.e. $c_s^2>0$ and $c_{GW}^2>0$. Following the reference \cite{Kobayashi:2019hrl}, the Lagrangian \eqref{lagrangian} can be understood as a sub-class of the Horndeski theory, where only the terms $G_2$ and $G_4$ of the Lagrangian (8) of \cite{Kobayashi:2019hrl} are non-zero. Given that, we can use their Eqs.~(42-45) joint with Eqs.~(49-53) to obtain the conditions under which our model does not present ghosts or gradient instabilities. When we apply this set of equations to the Lagrangian \eqref{lagrangian}, with the identifications $X\equiv - \frac{1}{2}\partial_{\mu}\phi\partial^{\mu}\phi$, $G_2(\phi,X)=X - \alpha\sqrt{2|X|}-V(\phi)$ and $G_4=\frac{1}{4}$, we obtain the following quantities:
\begin{eqnarray}
\mathcal{G}_T&=&2~G_4=\frac{1}{2},\\
\mathcal{F}_T&=&2~G_4=\frac{1}{2},\\
\Sigma&=&X G_{2X}+2X^2G_{2XX}-6H^2G_4=X-\frac{3}{2}H^2,\\
\Theta&=&2HG_4=\frac{H}{2},\\
\mathcal{G}_S&=&\frac{\Sigma}{\Theta^2}\mathcal{G}_T^2+3\mathcal{G}_T=\frac{X}{H^2},\\
\mathcal{F}_S&=&\frac{1}{a}\frac{d}{dt}\left(\frac{a}{\Theta}\mathcal{G}_T^2\right)-\mathcal{F}_T=-\frac{\dot{H}}{2H^2}.
\end{eqnarray}
Using the background equations \eqref{hd_a} and \eqref{hp_a} it is straightforward to obtain the propagation speeds of the tensor and scalar modes, given respectively by
\begin{eqnarray}
c_s^2&=&\frac{\mathcal{F}_S}{\mathcal{G}_S}=1+\frac{\alpha}{\sqrt{2X}},\\
c_{GW}^2&=&\frac{\mathcal{F}_T}{\mathcal{G}_T}=1.
\end{eqnarray}
Since $\alpha>0$ we guarantee that $c_s^2>0$ always. For the tensor speed $c_{GW}$ we also guarantee its positivity independent of the $\alpha$ value.

Lastly, in Sec. \ref{formalism_two}, we investigate the scenario of the cuscuton-like term in the presence of a second field, which we can interpret as an additional matter field. We need to ensure the stability of this additional field since matter perturbations are important for the late universe. In this case, the sound speed squared of the matter field reads as:
\begin{eqnarray}
c_m^2&=&\frac{P_Y}{P_Y+2YP_{YY}}=1>0
\end{eqnarray}
with the identifications $P(Y)=Y$, with $Y=\frac{1}{2}\partial_{\mu}\chi\partial^{\mu}\chi$ on Eq.~\eqref{lagrangian_two}. Adding a second field modifies the previous stability conditions for the cuscuton-like contribution. Now, we must look at its velocity propagation as given by $v^2=\left(\mathcal{F}_S-c^2_m Z\right)/\mathcal{G}_S$ with $\displaystyle Z=\left(\frac{\mathcal{G}_T}{\Theta}\right)^2\frac{(\rho+p)}{2c^2_m}$~\cite{Kobayashi:2019hrl}. Using $(\rho+p)$ evaluated for \eqref{rho_two} and \eqref{p_two} and the background equation \eqref{hd_two}, we show that $v^2=0$. This implies that the scalar modes are non-dynamical even inside the horizon, recovering the original cuscuton non-dynamical behavior.
Therefore, we ensure the absence of ghosts or gradient instabilities for all single or multi-field models, 
notwithstanding the cases with a phantom behavior for the equation of state, as we shall see in the next sections.
\section{First order framework for one field}\label{formalism}

Usually, the Hubble parameter $H$ is a function of time. In this work, we use the formalism developed in Ref. \cite{bglm2006}, in which $H$ is taken as a function $W$ of the scalar field, leading to first-order differential equations. We then take $H$ in the form
\be\label{hw}
    H = W(\phi).
\ee
This is a first-order differential equation for the scale factor. This assumption leads us to another first-order equation for the scalar field in the form
\be\label{phi_linha}
\dot{\phi} = -(W_{\phi} + \alpha),
\ee
where $W_{\phi} = dW/d\phi$. The solutions obtained from these first-order equations are also solutions of the second-order equation \eqref{eq_motion_2}. This approach is similar to the Hamilton-Jacobi formalism considered in Ref. \cite{ST} and in references therein. Using \eqref{hw} and \eqref{phi_linha} in \eqref{hd_a}, we obtain the following expression for the potential
\be\label{potential}
    V = \dfrac{3}{2}W^2 - \dfrac{1}{2}\left( W_{\phi} + \alpha \right)^2.
\ee
The expressions for energy density and pressure (\ref{rhoep}) can also be rewritten as follows, respectively
\begin{subequations}\label{rhoepw}
\be\label{rhow}
\rho = \dfrac{3}{2}W^2,
\ee
\be\label{pw}
p = W_{\phi} \left(W_{\phi} + \alpha \right) - \dfrac{3}{2}W^2.
\ee
\end{subequations}
Thus, the equation of state and the acceleration parameter are now given by
\be\label{omega_q_w}
    \dfrac{3}{2}(1 + \omega) \;=\; 1 - q \;=\; \dfrac{W_{\phi}}{W^2} \left(W_{\phi} + \alpha \right).
\ee
The above steps result in a first-order framework in the case of a single real scalar field. Thus, below we illustrate the main results by constructing two simple but interesting models of cosmic evolution driven by a quintessential scalar field in the presence of the additional cuscuton-like term.

\subsection{Single field with exponential potential}\label{one_exp}

As a first example we consider $W = Ae^{-B\phi}$, where $A$ and $B$ are real parameters. In this case, the potential \eqref{potential} is written as
\be\label{v_1}
V(\phi) = \dfrac{A^2}{2} \left(3 - B^2\right) e^{-2B\phi} + \alpha AB e^{-B\phi} - \dfrac{\alpha^2}{2}.
\ee
 In the case of standard dynamics ($\alpha=0$), one obtains an exponential potential of the form $V = V_0e^{-\lambda \phi}$. Cosmological scenarios driven by a scalar field with an exponential potential have been extensively studied in different contexts, including inflationary models \cite{halliwell1987, burd1988, wands1993, copeland1998} and theories of gravity such as scalar-tensor theories or string theories \cite{baumann2015, billyard}. A scalar field of this type is also an interesting candidate for dark energy \cite{huterer}, showing scaling solutions with a non-zero constant equation of state $\omega$ \cite{amendola2013, kim2013, bahamonde}. In this work, however, we want to show how the presence of the cuscuton-like contribution modifies cosmic evolution. In this context, the first-order equation \eqref{phi_linha} takes the form
\be\label{dotphi_1}
\dot{\phi} = AB e^{-B\phi} -\alpha.
\ee
The solution that simultaneously solves the first-order
equation \eqref{dotphi_1} and the second order equation \eqref{eq_motion_2} is as follows
\be\label{s_1}
\phi(t) = \dfrac{1}{B} \ln \left[ \dfrac{AB}{\alpha} \left(1 - e^{-\alpha B t} \right) \right].
\ee
The energy density and pressure are given by
\begin{subequations}\label{rhoep_1}
\be\label{rho_1}
\rho = \dfrac{3}{2} \left[ \dfrac{\alpha}{B \left( 1 - e^{-\alpha B t} \right)} \right]^2,
\ee
\be\label{p_1}
p = \left( B^2 - \dfrac{3}{2} \right) \left[ \dfrac{\alpha}{B \left( 1 - e^{-\alpha B t} \right)} \right]^2 - \dfrac{\alpha^2}{1- e^{-\alpha Bt}}.
\ee
\end{subequations}
The equation of state and the acceleration parameter are given by
\be\label{w_1}
\dfrac{3}{2}(1 + \omega) \;=\; 1 - q \;=\; B^2 e^{-\alpha Bt},
\ee
and the Hubble parameter has the form
\be\label{h_1}
H(t) = \dfrac{\alpha}{B \left( 1 - e^{-\alpha Bt} \right)}.
\ee
In this scenario, unlike the case $\alpha=0$, in which the equation of state is constant, we obtain an equation of state that is a function of time and that reaches the regime $\omega=-1$ when time grows enough, and the Hubble parameter then approaches a constant $\alpha/B$. 

\bfi[th!]
\bc
\includegraphics[scale=0.54]{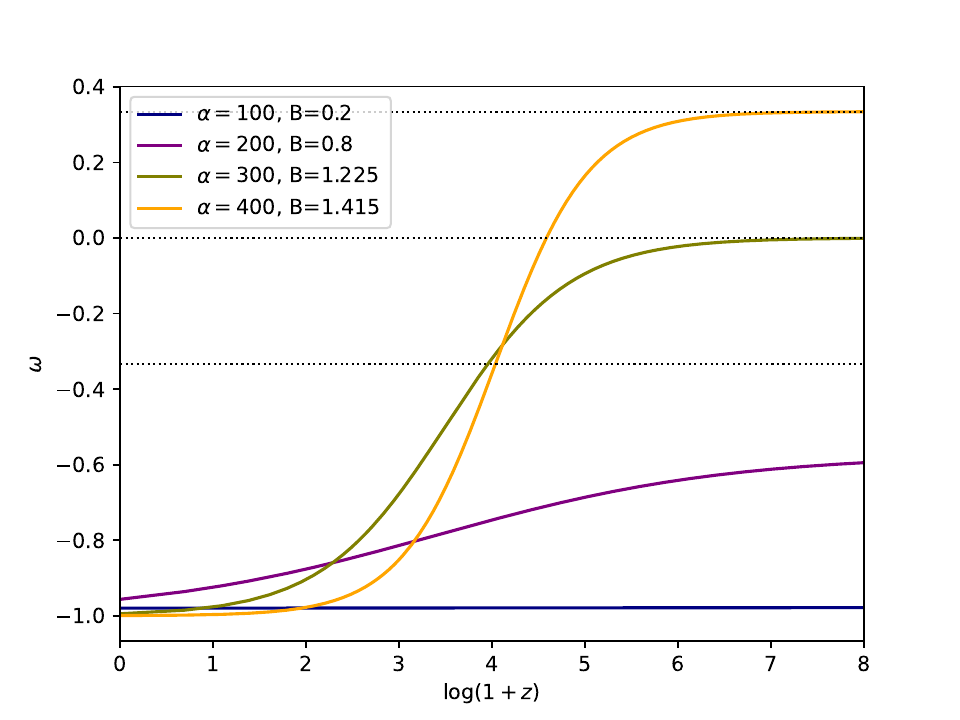}
\caption{Equation of state parameter \eqref{w_1} in function of the redshift, for different combinations of the parameters $\alpha$ and $B$. Note that we can have a transition from a radiation (or matter) dominated epoch to the current accelerated phase of the Universe depending on the values of $\alpha$ and $B$. Values of $B<1$ represent a dynamical dark energy component.}
\label{fig_walphaz}
\ec
\efi
%
%
%


In Cosmology, it is interesting to look at the evolution of a certain quantity with the redshift, $z$. To express the Hubble parameter in terms of $z$, remember that $H\equiv \dot{a}/a$ and the scale factor is related to the redshift as $a=1/(1+z)$. In this way, we can integrate Eq. \eqref{h_1} and invert it to get $t$ in function of $z$. Thus, we can rewrite the Hubble parameter as
\be
H(z) = \frac{\alpha \left[1+\left(1+z\right)^{B^2}\right]}{B}.
\label{Hz}
\ee
In the limit of $z=0$, we obtain $H_0=2\alpha/B$. Considering $H_0=70$\;km/s/Mpc and fixing $B=1$, for example, we obtain $\alpha=35$\;km/s/Mpc. As an alternative, we could consider measurements of different probes, such as the Hubble parameter, supernovae of type Ia and Baryonic Acoustic Oscillations (BAO), in order to put constraints on the model parameters. 

In Fig. \ref{fig_walphaz} we depict the behavior of the equation-of-state parameter \eqref{w_1} in function of the redshift for different values of $\alpha$ and $B$. We tested different combinations of $\alpha$ and $B$ and noticed that the upper limit of $B<1$ guarantees that $w$ is always below the line representing an accelerated universe ($w<-1/3$), and then going asymptotically to $-1$. Higher values of $B$ exhibit a transition from radiation ($B\sim 1.415$) or matter ($B\sim 1.225$) dominated phases to the accelerated one. Values lower than $0.5$ behave as a cosmological constant with $\omega\sim -1$. Due to the degeneracy between $\alpha$ and $B$, $w=-1$ can occur at a different time along the expansion history of the Universe. This behavior is similar to the one of a quintessence model, for which $w=-1$ today could represent the cosmological constant~(see \cite{Avsajanishvili:2023jcl} for a recent review on different classes of quintessence models).

\subsection{Single field with hyperbolic potential}\label{one_hyperb}

Our second class of potential is constructed from the choice $W = A \cosh (B\phi) - \alpha \phi$. This leads to the following potential
\be\label{v2}
V(\phi) = \dfrac{3}{2} \left[A \cosh(B\phi) - \alpha \phi \right]^2 - \dfrac{1}{2}A^2B^2 \sinh^2(B \phi).
\ee

\noindent This type of potential is similar to the one studied in Ref. \cite{afonso2006} in the context of bent brane, also constructed under the same first-order framework. In our model, we conveniently added a linear term $\alpha\phi$ to the W function. In this case, the first-order equation that we need to solve is written in the form
\be\label{dphi2}
\dot{\phi} = -AB \sinh(B \phi),
\ee

\noindent and its solution is
\be\label{p2}
\phi(t) = \dfrac{2}{B} \arctanh \left( e^{-AB^2t} \right).
\ee
As one can see, the cuscuton parameter does not interfere with the evolution of the field $\phi$, but it does change the behavior of the equation of state compared to the standard dynamics. The equation of state and the acceleration parameter are given by

\begin{widetext}
\be\label{w2}
\dfrac{3}{2}(1+ \omega) \;=\; 1 - q \;=\; \dfrac{ AB^3 \left[ AB \sinh \left( 2 \arctanh ( e^{-AB^2t} ) - \alpha \right) \right] \sinh \left( 2 \arctanh ( e^{-AB^2t} ) \right) }{ \left[ AB \cosh \left( 2 \arctanh ( e^{-AB^2t} ) \right) - 2\alpha \arctanh ( e^{-AB^2t} ) \right]^2}.
\ee
\end{widetext}

\bfi[th]
\bc
\includegraphics[scale=0.6]{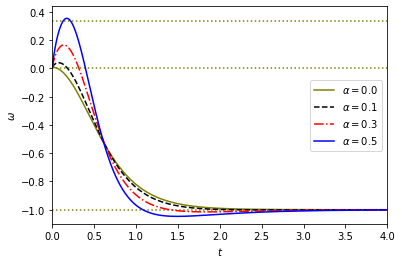}
\caption{Equation of state \eqref{w2} for $\alpha = 0$ (olive solid), $0.1$ (black dashed), $0.3$ (red dash-dotted) and $0.5$ (blue solid), with $A = 1$ and $B = 1.23$. The horizontal dotted lines, from top to bottom, indicate that $\omega=1/3$ (radiation), $0$ (matter), and $-1$ (cosmological constant).}
\label{fig_walpha2}
\ec
\efi

The $\omega$ in equation \eqref{w2} is depicted in Fig. \ref{fig_walpha2} for several distinct values of $\alpha$. We can see that the presence of the term $\alpha$ entails different possibilities for cosmic evolution. For the choice of parameters $A=1$ and $B=1.23$, the olive solid curve ($\alpha=0$) indicates a transition from the matter-dominated phase directly to the cosmological constant phase. For higher values of $\alpha$, however, there are other possibilities of evolution. The blue solid curve, for example, represents an evolution that transits between the phases of domination of matter and radiation and goes towards the $\omega=-1$ regime, passing through a phantom phase. The study of cosmological models allowing the phantom scenario is of current relevance \cite{caldwell2002, caldwell2003, carroll2003, faraoni2005, tripathy2020}, because we may be living in a phantom stage which is not ruled out by the current observational data \cite{aghanim2020}. As discussed in \cite{caldwell2002}, an equation of state $\omega<-1$ cannot be achieved with Einstein gravity and a canonical Lagrangian. Therefore, the current example describes a cosmic evolution in which the existence of the phantom phase is conditioned by the presence of the cuscuton term, with $\alpha\neq0$.

Using \eqref{p2} in the expression for $W$ of this model, we obtain the Hubble parameter as being given by
\be\label{h_2}
\begin{split}
    H(t) = &\; A \cosh\left[ 2 \arctanh\left( e^{-AB^2t} \right) \right]\\
    &\; -\dfrac{2\alpha}{B} \arctanh\left( e^{-AB^2t} \right).
\end{split}
\ee
To get it in terms of $z$, we can proceed as we did at the end of Sec. III.A.

\section{First order framework for two fields}\label{formalism_two}

In this Section, we study other interesting cosmological scenarios, but now we make the field with the additional cuscuton-like contribution to interact with another scalar field, which has standard dynamics. We therefore need to extend the formalism developed previously to the case of two interacting fields.
The new Lagrangian is of the form
\be\label{lagrangian_two}
{\cal L} = - \dfrac{1}{2}\partial_{\mu}\phi \partial^{\mu}\phi -\alpha \sqrt{|\partial_{\mu}\phi \partial^{\mu}\phi|} - \dfrac{1}{2}\partial_{\mu}\chi\partial^{\mu}\chi - V(\phi,\chi).
\ee

\noindent The fields obey the following equations of motion
\begin{subequations}\label{eq_motion_two}
\be
\ddot{\phi} + 3H(\dot{\phi} + \alpha) + V_{\phi} = 0,
\ee
\be
\ddot{\chi} + 3H\dot{\chi} + V_{\chi} = 0,
\ee
\end{subequations}
where $V_{\phi}=\partial V/\partial\phi$, $V_{\chi}=\partial V/\partial\chi$, and the energy density and pressure are
\begin{subequations}\label{rhoep_two}
\begin{equation}\label{rho_two}
    \rho = \frac{1}{2}\dot{\phi}^2 + \frac{1}{2}\dot{\chi}^2 + V(\phi,\chi),
\end{equation}
\begin{equation}\label{p_two}
    p = \frac{1}{2}\dot{\phi}^2 + \alpha\; \dot{\phi}\; + \frac{1}{2}\dot{\chi}^2 - V(\phi,\chi).
\end{equation}
\end{subequations}

\noindent The Friedmann equations in this case are written as
\begin{subequations}\label{frid_two}
\begin{equation}\label{hd_two}
    3H^2 = \dot{\phi}^2 + \dot{\chi}^2 + 2V,
\end{equation}
\begin{equation}\label{hp_two}
    \dot{H} = - (\dot{\phi} + \alpha )\dot{\phi} - \dot{\chi}^2.
\end{equation}
\end{subequations}
To build the first-order formalism, we suppose the Hubble parameter is a function of the two fields. Here we follow the suggestions of Refs. \cite{bglm2006,bbl2006}, and we consider $W$ an additive function, so that we have
\be
H = W_1(\phi) + W_2(\chi).
\ee
As before, this is a first-order equation for $a$, which allows us to write the following first-order equations for the fields
\begin{subequations}\label{phi_linha_two}
\be
\dot{\phi} = -(W_{\phi} + \alpha),
\ee
\be
\dot{\chi} = -W_{\chi}.
\ee
\end{subequations}
The potential now takes the form
\be\label{potential_two}
    V = V_1 + V_2 + V_{\rm int},
\ee
where $V_1= 3W_1^2/2 - (W_\phi + \alpha)^2/2$, $V_2= 3W_2^2/2 - W_\chi^2/2$ and $V_{\rm int} = 3W_1W_2$, with $V_{\rm int}$ representing the interaction between the two fields. The energy density and pressure can also be rewritten as follows
\begin{subequations}\label{rhoepw_two}
\be\label{rhow}
\rho = \dfrac{3}{2}\left(W_1 + W_2\right)^2,
\ee
\be\label{pw}
p = W_{\phi} \left(W_{\phi} + \alpha \right) + W_{\chi}^2 - \dfrac{3}{2}\left(W_1 + W_2\right)^2.
\ee
\end{subequations}

\noindent The equation of state and acceleration parameter is then given by
\be\label{omega_q_w_two}
    \dfrac{3}{2}(1 + \omega) \;=\; 1 - q \;=\; \dfrac{W_{\phi}\left(W_{\phi} + \alpha \right) + W_{\chi}^2}{(W_1 + W_2)^2}.
\ee

One way to think about the relative presence of each energetic component in the universe is to define its density in relation to the total density. We therefore define the relative density of each component as $\Omega_i=W_i^2/W^2$. From \eqref{rhow} we can then write
\be
\Omega_1 + \Omega_2 + \Omega_{\rm int} = 1,
\ee
where $\Omega_1=W_1^2/(W_1+W_2)^2$, $\Omega_2=W_2^2/(W_1+W_2)^2$ and $\Omega_{\rm int}=(2W_1W_2)/(W_1+W_2)^2$. Here we have separated the interaction relative density of the pure densities of each field, and below we study some quintessential two-field models with the presence of the additional cuscuton contribution.

\subsection{Two fields with exponential potential}\label{two_exp}

We were able to build an interesting model that describes the transition between an arbitrary phase of cosmic evolution and the cosmological constant phase by making the choice $W = Ae^{-B\phi} + Ce^{-D\chi}$. In this case, the potential has the form
\be
\begin{split}
V(\phi,\chi) = & \; \dfrac{A^2}{2}\left(3 - B^2\right)e^{-2B\phi} + \dfrac{C^2}{2}\left(3 - D^2\right)e^{-2D\chi}\\[0.3cm]
& + A\left(\alpha B + 3Ce^{-D\chi} \right)e^{-B\phi} - \dfrac{\alpha^2}{2}.
\end{split}
\ee
The solutions of equations \eqref{eq_motion_two} are
\begin{subequations}
\be
\phi(t) = \dfrac{1}{B} \ln\left[ \dfrac{AB}{\alpha} \left( 1 - e^{-\alpha B t} \right) \right],
\ee
\be
\chi(t) = \dfrac{1}{D} \ln\left( CD^2 t \right).
\ee
\end{subequations}
The equation of state and acceleration parameter are given by
\be\label{wq_two}
\dfrac{3}{2}(1+\omega) = 1 - q = \dfrac{ B^2D^4 \left[ \alpha^2 t^2 e^{-\alpha Bt} + \left(1 - e^{-\alpha Bt}\right)^2 \right] }{ \left[ \alpha D^2t + B \left(1 - e^{-\alpha Bt}\right) \right]^2 },
\ee
and the relative energy densities are
\begin{subequations}
\be
\Omega_1 = \dfrac{\alpha^2 D^4 t^2}{ \left[ \alpha D^2t + B ( 1 - e^{-\alpha Bt}) \right]^2 },
\ee
\be
\Omega_2 = \dfrac{B^2D^2(1 - e^{-\alpha Bt})^2}{ \left[ \alpha D^2t + B \left( 1 - e^{-\alpha Bt} \right) \right]^2 },
\ee
\be
\Omega_{\rm int} = \dfrac{ 2\alpha BD^2 t (1 - e^{-\alpha Bt}) }{ \left[ \alpha D^2t + B \left( 1 - e^{-\alpha Bt} \right) \right]^2 }.
\ee
\end{subequations}

\bfi[h!]
\bc
\includegraphics[scale=0.54]{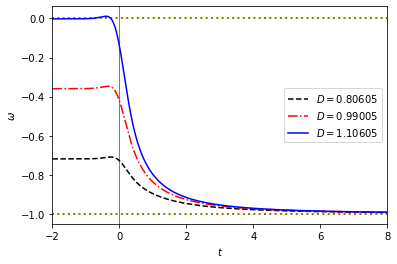}
\includegraphics[scale=0.54]{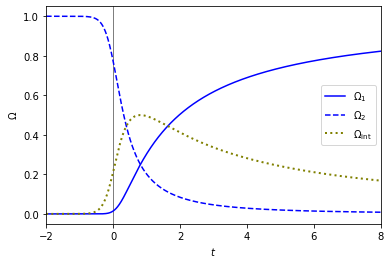}
\caption{Upper panel shows the equation of state \eqref{wq_two} for $D = 0.80605$ (black dashed), $0.99005$ (red dash-dotted) and $1.10605$ (blue solid), with $\alpha = 3$ and $B = 0.05$. The lower panel shows the relative densities of fields and interactions corresponding to the solid blue line in the upper panel.}
\label{fig_wd}
\ec
\efi

\bfi[h!]
\bc
\includegraphics[scale=0.54]{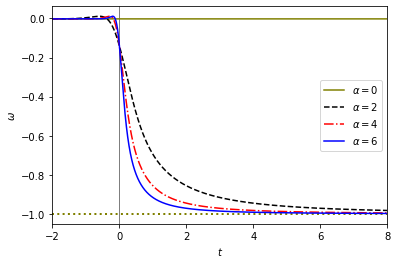}
\includegraphics[scale=0.54]{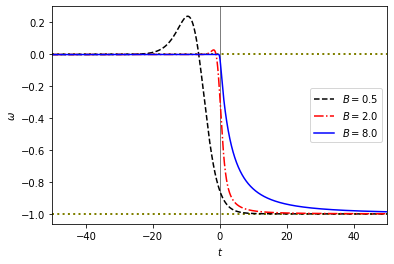}
\caption{Upper panel shows the equation of state \eqref{wq_two} for $\alpha = 0$ (olive solid), $2$ (black dashed), $4$ (red dash-dotted) and $6$ (blue solid), with $D = 1.10605$ and $B = 0.05$. Lower panel shows the equation of state for $B = 0.5$ (black dashed), $2$ (red dash-dotted) and $8$ (blue solid), with $D = 1.10605$ and $\alpha = 1$.}
\label{fig_walpha}
\ec
\efi

The behavior of $\omega$ is shown in Figs. \ref{fig_wd} and \ref{fig_walpha}. The $D$ parameter controls the value of $\omega$ at the beginning of the first phase of the considered evolution. In Fig. \ref{fig_wd} we see that, for our choice of parameter values, the first phase of the blue solid curve corresponds to the predominance of matter. However, different initial conditions lead to the same final configuration: the $\omega=-1$ regime. In the lower panel of the same Figure, we see that the relative interaction density $\Omega_{\rm int}$ becomes important during the transition between phases and decreases again when the field with additional cuscuton term becomes predominant.

The parameter $\alpha$, as can be seen in Fig. \ref{fig_walpha}, controls the start and end of the transition between phases. It therefore controls the duration of this transition. Larger values of $\alpha$ correspond to faster transitions. It is important to realize that if $\alpha=0$, in which case the cuscuton term does not exist, the transition between the phases does not occur because the equation of state is constant (olive solid curve). It is still possible to think of a fluctuation around the value of $\omega$ in the initial phase before the start of the transition, and this is controlled by the parameter $B$. In the lower panel of Fig. \ref{fig_walpha}, we see that the greater the value of $B$, the smaller the amplitude of this fluctuation. The blue solid curve, for example, indicates a transition where this fluctuation practically does not appear. The cosmic evolution starts from the domination phase of matter and the equation of state decreases until it reaches the behavior of a cosmological constant.

The Hubble parameter of this model has the form
\be\label{h_3}
H(t) = \dfrac{\alpha}{B (1 - e^{-\alpha Bt})} + \dfrac{1}{D^2t}.
\ee
To get it in terms of $z$, we can proceed as we did at the end of Sec. III.A.

\subsection{Two fields with hyperbolic potential}\label{two_cosh}

The new model is now built from the choice $W = A\cosh(B\phi) - \alpha\phi + C\cosh(D\chi)$, which also allows us to study transitions between an arbitrary phase of cosmic evolution and the cosmological constant phase. This leads to potential
\be\label{v4}
\begin{split}
V(\phi,\chi) = & \;  \dfrac{3}{2} \left[A \cosh(B\phi) - \alpha \phi \right]^2 - \dfrac{A^2B^2}{2} \sinh^2(B \phi)\\[0.2cm]
& + \dfrac{ C^2}{2} \left[3 \cosh^2(D\chi) - D^2 \sinh^2(D \chi) \right]\\[0.2cm]
& + 3C \left[ A\cosh(B\phi) - \alpha\phi \right] \cosh(D\chi).
\end{split}
\ee

\noindent The solutions of the equations \eqref{eq_motion_two} are
\begin{subequations}
\be
\phi(t) = \dfrac{2}{B} \arctanh \left(e^{-AB^2t}\right),
\ee
\be
\chi(t) = \dfrac{2}{D} \arctanh \left(e^{-CD^2t}\right).
\ee
\end{subequations}
To help us write the results, from now on we will introduce the functions
\begin{subequations}
\be
G_1(t)\equiv \arctanh (e^{-AB^2 t}),
\ee
\be
G_2(t)\equiv \arctanh(e^{-CD^2 t})
\ee
\end{subequations}
\noindent The equation of state, the acceleration parameter, and the relative energy densities are given by

\begin{widetext}

\be\label{w4}
\begin{split}
\dfrac{3}{2}(1+\omega)  = 1 - q
 = \dfrac{AB \left[ AB \sinh\left( 2G_1(t) \right) - \alpha \right]\;\sinh\left( 2G_1(t) \right) + C^2D^2 \sinh^2\left( 2G_2(t) \right) }{ \left[ A\cosh\left( 2G_1(t) \right) + C\cosh\left( 2G_2(t) \right) -\frac{2\alpha}{B}G_1(t) \right]^2 }
\end{split}
\ee

\begin{subequations}\label{densities_a}
\be
\Omega_1 = \dfrac{ \left[ A\cosh\left( 2G_1(t) \right) - \frac{2\alpha}{B} G_1(t) \right]^2 }{ \left[ A\cosh\left( 2G_1(t) \right) + C\cosh\left( 2G_2(t) \right) -\frac{2\alpha}{B}G_1(t) \right]^2 }
\ee
\be
\Omega_2 = \dfrac{ C^2 \cosh^2\left( 2 G_2(t)\right) }{ \left[ A\cosh\left( 2G_1(t) \right) + C\cosh\left( 2G_2(t) \right) -\frac{2\alpha}{B}G_1(t) \right]^2 }
\ee
\be
\Omega_{\rm int} = \dfrac{ 2C \left[ A\cosh\left( 2G_1(t) \right) -\frac{2\alpha}{B}G_1(t) \right] \cosh\left( 2G_2(t) \right) }{ \left[ A\cosh\left( 2G_1(t) \right) + C\cosh\left( 2G_2(t)\right) -\frac{2\alpha}{B}G_1(t) \right]^2 }
\ee
\end{subequations}

\end{widetext}

\bfi[h!]
\bc
\includegraphics[scale=0.54]{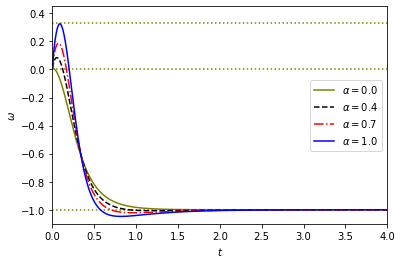}
\includegraphics[scale=0.54]{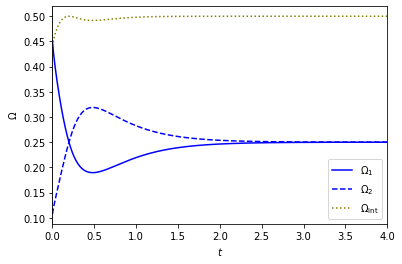}
\caption{Upper panel shows the equation of state \eqref{w4} for $\alpha = 0$ (olive solid), $0.4$ (black dashed), $0.7$ (red dash-dotted) and $1.0$ (blue solid), with $A = 1$, $B = 1.5$, $C = 1$ and $D = 2.15$. The horizontal dotted lines, from top to bottom, respectively indicate $\omega=1/3$ (radiation), $0$ (matter), and $-1$ (cosmological constant). The lower panel shows the relative
densities of fields and interactions corresponding to the blue
solid line in the upper panel.}
\label{fig_walpha3}
\ec
\efi

The equation \eqref{w4} is displayed in the upper panel of Fig. \ref{fig_walpha3}. For the choice of parameters, all curves indicate cosmic evolutions starting from a characteristic phase of matter predominance. In the absence of the cuscuton contribution ($\alpha=0$), the transition between the phases $\omega=0$ and $\omega=-1$ occurs directly, that is, there is no fluctuation around $\omega= 0$ before $\omega$ decreases and not a phantom phase before the $\omega=-1$ regime is reached. When $\alpha\neq 0$, however, we see that both the fluctuation around $\omega=0$ and the phantom phase exist. In particular, the solid blue curve describes a cosmic evolution in which the initial phase fluctuates between matter and radiation predominance and passes through a significant phantom phase before reaching the cosmological constant phase. The lower panel shows that the relative interaction density $\Omega_{\rm int}$ is practically constant throughout evolution and predominates over other densities. The relative density of the field with additional cuscuton term $\Omega_1$ predominates at the beginning over the density of the field with standard dynamics $\Omega_2$. There is a new phase in which $\Omega_2$ becomes predominant over $\Omega_1$, and then the $\omega=-1$ regime is reached when $\Omega_1$ and $\Omega_2$ equal each other and remain the same.

In this model, the Hubble parameter takes the form
\be
\begin{split}
    H(t) = &\; A\cosh\left[ 2\arctanh\left( e^{-AB^2t} \right) \right]\\
    &\; + C\cosh\left[ 2\arctanh\left( e^{-CD^2t} \right) \right]\\
    &\; -\dfrac{2\alpha}{B}\arctanh\left( e^{-AB^2t} \right).
\end{split}
\ee
To get it in terms of $z$, we can proceed as we did at the end of Sec. III.A.

\subsection{Combining exponential and hyperbolic potentials}\label{two_cosh}

As a first way to combine the two types of potential from the previous sections, we consider the field  $\phi$ related to the exponential potential and $\chi$ related to the hyperbolic potential, so that $W = Ae^{-B\phi} + C\cosh(D\chi)$. In this case, the potential has the form
\be\label{v5}
\begin{split}
V(\phi,\chi) = & \;  \dfrac{A^2}{2} \left(3 - B^2\right)e^{-2B\phi}\\[0.3cm]
& + \dfrac{C^2}{2}\left[3\cosh^2(D\chi) - D^2\sinh^2(D\chi)\right]\\[0.3cm]
& + A\left[\alpha B + 3C \cosh(D\chi)\right]e^{-B\phi} - \dfrac{\alpha^2}{2}.
\end{split}
\ee

\noindent The solutions of the equation \eqref{eq_motion_two} are
\begin{subequations}
\be
\phi(t) = \dfrac{1}{B} \ln \left[\dfrac{AB}{\alpha} \left(1 - e^{-\alpha Bt}\right) \right],
\ee
\be
\chi(t) = \dfrac{2}{D} \arctanh \left(e^{-CD^2t}\right)=\dfrac{2}{D}G_2(t),
\ee
\end{subequations}

\noindent and the equation of state and the acceleration parameter are given by

\ben\label{w5}
\dfrac{3}{2}(1+\omega) = 1-q=\qquad\qquad\qquad\qquad\qquad\qquad\qquad\nonumber\\
\!=\! \dfrac{ \alpha^2e^{-\alpha Bt} + C^2D^2 (1\!-e^{-\alpha Bt})^2 \sinh^2\left(2G_2(t) \right) }{ \left[ \frac{\alpha}{B}\! +\! C(1\!-\!e^{\alpha Bt}) \cosh\left(2 G_2(t) \right) \right]^2 }.\;\;\;
\een
The relative energy densities are as follows
\begin{subequations}\label{densities_b}
\be
\Omega_1 = \dfrac{\alpha^2}{ \left[ \alpha + BC(1-e^{\alpha Bt}) \cosh\left(2 G_2(t)\right) \right]^2  },
\ee
\be
\Omega_2 = \dfrac{ B^2C^2 (1-e^{-\alpha Bt})^2 \cosh^2\left(2G_2(t)\right) }{ \left[ \alpha + BC(1-e^{\alpha Bt}) \cosh\left(2 G_2(t)\right) \right]^2  },
\ee
\be
\Omega_{\rm int} = \dfrac{ 2\alpha BC (1-e^{-\alpha Bt})  \cosh \left(2 G_2(t)\right) }{ \left[ \alpha + BC(1-e^{\alpha Bt}) \cosh\left(2 G_2(t)\right) \right]^2  }.
\ee
\end{subequations}
The equation \eqref{w5} is depicted in the upper panel of Fig. \ref{fig_walpha4} for several values of $\alpha$. In this scenario, we see that for fixed $D$, the $\alpha$ parameter controls the starting value of the equation of state. For our choice of values for parameters $B$, $C$ and D, $\alpha = 0$ corresponds to a beginning of evolution approximately in the phase of matter, while $\alpha=4.4$ corresponds to a beginning of evolution approximately in the phase of radiation. The late evolution, however, is the same for all different initial conditions. It is possible to notice that, in this scenario, the cuscuton term does not contribute to a fluctuation around the initial phase nor the existence of a phantom phase. The parameters $B$ and $D$ also control the initial conditions. The parameter $C$, on the other hand, does not interfere with the initial conditions, but controls the slope of the curves along the evolution, as can be seen in the lower panel of Fig \ref{fig_walpha4}. The higher the value of $C$, the faster the transition between phases.

\bfi[b]
\bc
\includegraphics[scale=0.54]{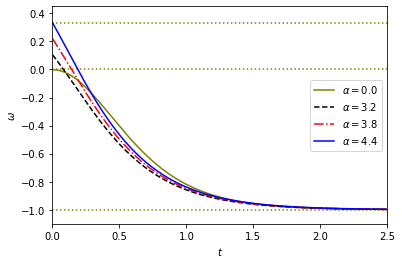}
\includegraphics[scale=0.54]{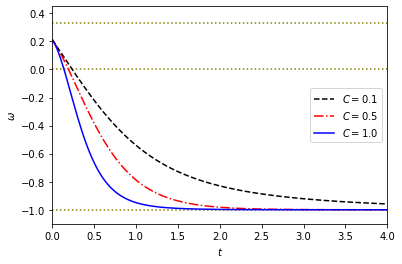}
\caption{Equation of state \eqref{w5} (upper panel) for $\alpha = 0$ (olive solid), $3.2$ (black dashed), $3.8$ (red dash-dotted) and $4.4$ (blue solid), with $A=1$, $B=2.5$, $C=1$ and $D=1.224$; and (Lower panel) for $C=0.1$, $0.5$ and $1.0$, with $\alpha = 1$, $A=1$, $B=2.5$ and $D=1.6$.}
\label{fig_walpha4}
\ec
\efi

\bfi[ht!]
\includegraphics[scale=0.54]{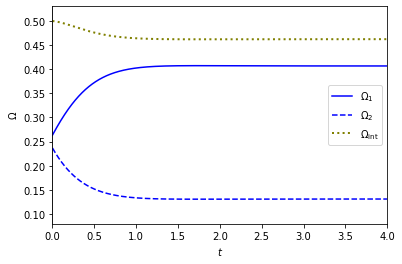}
\caption{Relative densities \eqref{densities_b} of fields and interaction corresponding to the blue solid line in the upper panel of Fig. \ref{fig_walpha4}.}
\label{fig_ow4}
\efi

The relative densities \eqref{densities_b} are shown in Fig. \ref{fig_ow4}. For our choice of parameter values, we have a scenario where the relative density of the field with additional cuscuton term predominates over the density of the field with standard dynamics, and in which the interaction density predominates over the others, but this is not the only possible scenario.

The Hubble parameter in this case is written as
\be
\begin{split}
    H(t) = &\; \dfrac{\alpha}{B (1-e^{-\alpha Bt})}\\
    &\; + C \cosh\left[ 2 \arctanh\left( e^{-CD^2t} \right) \right].
\end{split}
\ee
To get it in terms of $z$, we can proceed as we did at the end of Sec. III.A.

Another possibility to combine the potentials of the previous sections into a single model is to consider $W = A\cosh(B\phi) - \alpha \phi + Ce^{-D\chi}$. The field solutions in this case are
\begin{subequations}
\be
\phi(t) = \dfrac{2}{B} \arctanh \left(e^{-AB^2t}\right)=\dfrac{2}{B}G_1(t),
\ee
\be
\chi(t) = \dfrac{1}{D} \ln \left(CD^2 t \right).
\ee
\end{subequations}
The equation of state and the acceleration parameter are now given by

\ben\label{w6}
\dfrac{3}{2}(1+\omega) = 1-q=\qquad\qquad\qquad\qquad\qquad\qquad\qquad\nonumber\\
\!=\! \dfrac{ At^2 \left[ a\sinh\left( 2G_1(t) \right) - \alpha \right] \sinh\left( 2G_1(t) \right) + \frac{1}{AD^2} }{ \left[ At \cosh\left( 2G_1(t) \right)- \frac{2\alpha t}{B} G_1(t) + \frac{1}{D^2} \right]^2 },\;\;\;
\een

\noindent and the relative densities become
\begin{subequations}\label{densities_5}
\be
\Omega_1 = \dfrac{ D^4t^2 \left[ AB\cosh\left(2G_1(t) \right) - 2\alpha G_1(t) \right]^2 }{ \left[ ABD^2t \cosh\left(2G_1(t) \right) - 2D^2\alpha t G_1(t)+ B \right]^2 },
\ee
\be
\Omega_2 = \dfrac{ B^2 }{ \left[ ABD^2t \cosh\left(2 G_1(t) \right) - 2D^2\alpha t G_1(t)+ B \right]^2 },
\ee
\be
\Omega_{\rm int} = \dfrac{ 2BD^2t \left[ AB\cosh\left(2G_1(t) \right) - 2\alpha G_1(t) \right] }{ \left[ ABD^2t \cosh\left(2G_1(t) \right) - 2D^2\alpha t G_1(t) + B\right]^2 }.
\ee
\end{subequations}

\bfi[h!]
\includegraphics[scale=0.54]{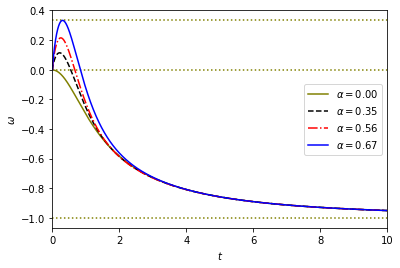}
\includegraphics[scale=0.54]{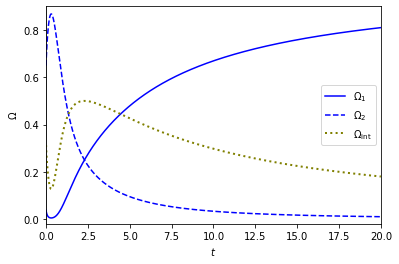}
\caption{Upper panel shows the equation of state \eqref{w6} for $\alpha = 0$ (olive solid), $0.35$ (black dashed), $0.56$ (red dash-dotted) and $0.67$ (blue solid), with $A = 0.2$, $B = 3.05$, $C = 1$ and $D = 1.5$. The horizontal dotted lines, from top to bottom, respectively indicate $\omega=1/3$ (radiation), $0$ (matter), and $-1$ (cosmological constant). The lower panel shows the relative
densities of fields and interactions corresponding to the blue
solid line in the upper panel.}
\label{fig_walpha5}
\efi

The equations \eqref{w6} and \eqref{densities_5} are displayed in Fig. \ref{fig_walpha5}. This time, the $\alpha$ parameter does not control the starting value of $\omega$, as in Fig. \ref{fig_walpha4}, but the fluctuation around the initial phase, similar to the case in Fig. \ref{fig_walpha3}. The higher the value of $\alpha$, the greater the variation around the initial value of $\omega$, but all different initial conditions lead to the same cosmic evolution after some time. In all cases, the $\omega=-1$ regime is reached without the existence of a previous phantom phase. In the lower panel of Fig. \ref{fig_walpha5}, we see that the relative density $\Omega_2$ predominates initially but then starts to decrease, while the density $\Omega_1$ increases. The initially small interaction density $\Omega_{\rm int}$ starts to increase and reaches its maximum during the transition between $\Omega_1$ and $\Omega_2$. The interaction then slows down and, during the rest of the evolution, $\Omega_1$ predominates.

The Hubble parameter here is
\be
\begin{split}
    H(t) = &\; A\cosh\left[ 2\arctanh\left( e^{-AB^2t} \right) \right]\\
    &\; -\dfrac{2\alpha}{B}\arctanh\left( e^{-AB^2t} \right) + \dfrac{1}{D^2t}. 
\end{split}
\ee
To get it in terms of $z$, we can proceed as we did at the end of Sec. III.A.

\section{Observational constraints: combining exponential and matter fields}
\label{observa}

Let's consider the scenario in which the action is given by
\be\label{actionf}
S_T=\int \mathrm{d}^4x \sqrt{-g} \left(\frac{1}{4} R + \mathcal{L}(\phi,\partial_{\mu}\phi) + \mathcal{L}_{m} \right),
\ee
where $\mathcal{L}_{m}$ represents the standard matter-energy Lagrangian and $\mathcal{L}(\phi,\partial_{\mu}\phi)$ stands for the Lagrangian \eqref{lagrangian}. 
When varying the action \eqref{actionf} with respect to the metric and considering the energy-momentum tensor of a perfect fluid, the Friedmann equation now reads as
\be \label{Hnew}
3H^2 = 2\rho_T,
\ee
where $\rho_T$ stands for the total density. With the universe filled with the scalar field $\phi$, whose energy density we call $\rho_{CL}$ standing for cuscuton-like, plus matter (baryonic and dark matter) we have
\be\label{landl}
\rho_T = \rho_{CL} + \rho_m.
\ee
Inspired by the results found in Sec.III-A, we suppose a parametrization for the energy density $\rho_{CL}$ as follows
\be
\rho_{CL}=3\left[\frac{\alpha \left(1+\left(1+z\right)^{B^2}\right)}{B}\right]^2.
\ee
%
Note that in the limit of $z=0$, we have $\rho_{CL,0}=12(\alpha/B)^2$, which behaves as a cosmological constant.

It is useful to rewrite Eq.~\eqref{Hnew} as
\be\label{constraint}
1=2\frac{(\rho_{CL} + \rho_m)}{3H^2} = \Omega_{CL} + \Omega_m
\ee
where we have used the critical density definition $\rho_c=3H^2/2$ and $\Omega_i\equiv \rho_i/\rho_c$. This constraint yield a direct relation between $\rho_{CL,0}$ and $\Omega_{m,0}$, resulting in the following expansion history parameter, $H(z)$,
\be \label{hphimat}
\frac{H(z)^2}{H_0^2}=\Omega_{m,0}(1+z)^3 + (1-\Omega_{m,0})\frac{\left[1+(1+z)^{B^2}\right]^2}{4},
\ee
where $\Omega_{m,0}$ is the matter density parameter today. It is important to notice that in the limit of $z=0$, we recover the standard $\Lambda$CDM model with $\omega=-1$. This approach is similar to parameterizing $\omega$ as a function of $z$, an often-used linear redshift parameterization of the equation of state, where we recover $\omega=-1$ for $z=0$.

At this point, we are able to use observational probes to put constraints on the model parameters by performing a Monte Carlo Markov Chain analysis, for example. We decided to use the SimpleMC code~\cite{BOSS:2014hhw,simplemc}, which is an MCMC code for cosmological parameter estimation where only the expansion history of the Universe is accounted for. The code solves the cosmological equations for the background similar to CLASS or CAMB codes, but it is much faster since it does not do perturbations calculations and uses only a geometric CMB compressed likelihood. We implemented a new module, where the background is calculated based on Eq. \eqref{hphimat}, and considered as free parameters (with flat priors): the matter density parameter $\Omega_{m,0} \in [0.1,0.5]$, the dimensionless Hubble constant $h = H_0/100 \in [0.4,1]$, and the dimensionless model parameter $B \in [0.001,1]$. In addition, we also used a BBN prior to the baryon physical density $\Omega_b h^2$ according to \cite{Cooke:2013cba}. Finally, the $\alpha$ parameter can be derived using \eqref{constraint}, as 
\be
\alpha=\frac{BH_0}{\sqrt{8}}\sqrt{1-\Omega_{m,0}}.
\ee

\begin{figure*}[ht!]
    \centering
    \includegraphics[width=\textwidth]{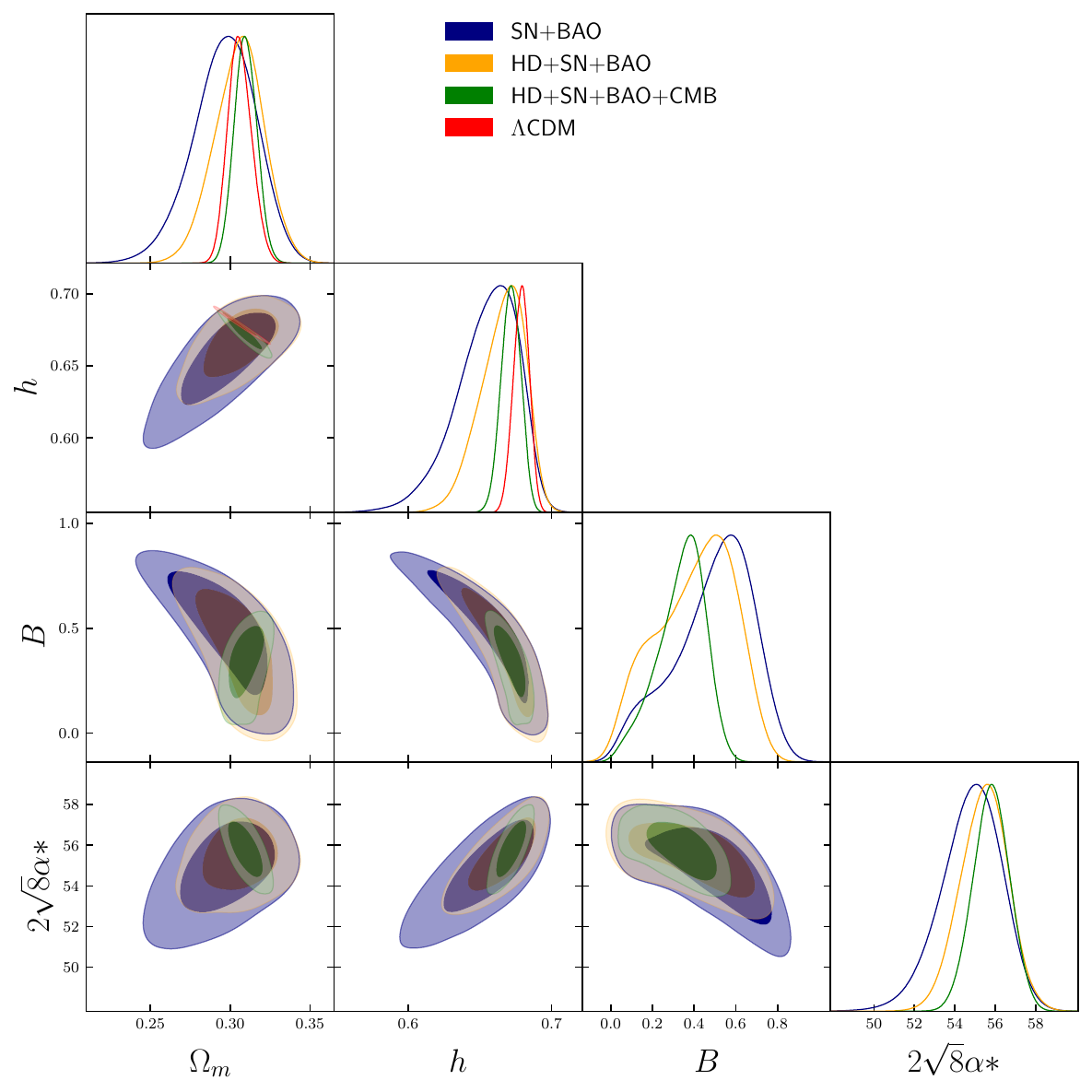}
    \caption{One-dimensional posterior distributions and two-dimensional joint contours for the parameter space $\{\Omega_m, h, B\}$ for different dataset combinations. We display also the $\Lambda$CDM model, using the full dataset HD+SN+BAO+CMB, for comparison purposes.}
    \label{fig:tri_LCDM+cuscuton}
\end{figure*}


We used different dataset combinations, including Cosmic Chronometers measurements~\cite{Moresco:2016mzx}, type Ia supernovae from Pantheon+ compilation (SN)~\cite{Brout:2022vxf}, Baryonic Acoustic Oscillations (BAO)~\cite{BOSS:2016wmc,eBOSS:2017cqx,eBOSS:2019ytm,eBOSS:2019qwo,Beutler,Ross:2014qpa}, and geometric CMB data, as given by the ratio of the comoving angular diameter distance to the sound horizon~\cite{Planck:2015fie}. The results are presented in Table \ref{tab:Tabel_results_1} and Fig.~\ref{fig:tri_LCDM+cuscuton}. The constraints for the matter density parameters (either only baryonic or baryonic+dark-matter components) present a good agreement with the $\Lambda$CDM model. The inclusion of the $B$ parameter affects the evolution of the e.o.s parameter $\omega$ and, consequently, the derived values for $H_0$. The mean values of $H_0$ are slightly lower than the $\Lambda$CDM ones, but the big uncertainties ensure its agreement within $1\sigma$ C.L. with the standard model. We can not say the same regarding $\omega$, which excludes $\omega=-1$ within $1\sigma$ when considering only Cosmic Chronometers data. On the other side, when extending the dataset to include SN, BAO, and CMB data, we recover the cosmological constant behavior for the e.o.s parameter and obtain very good constraints for the model parameters $B$ and $\alpha$. The goodness of the cuscuton-like model is therefore attested by these results and also when we perform a model comparison test using the Akaike Information Criteria (AIC)~\cite{AIC}. 

The SimpleMC code also provides the AIC values, which consider the point that maximizes the posterior probability distribution to compare the models, taking into account the number of extra parameters of the model under consideration. We can adopt a scale based on the difference between the model we want to test and the one providing the minimum AIC value: $\Delta AIC = AIC_i - AIC_{min}$, such that models with $\Delta AIC\leq 2$ have substantial support, models for which $4<\Delta AIC <7$ have less support, and those with $\Delta AIC>10$ have no support at all compared to the best-fit model~\cite{AICscale}. We chose $\Lambda$CDM as the reference model and calculated the AIC value for each dataset combination considered. The cuscuton-like model is moderately disfavored if we consider only the Cosmic Chronometers data, with $\Delta$AIC$=4.1$. The dataset combinations SN+BAO, HD+SN+BAO, and HD+SN+BAO+CMB have $\Delta$AIC$=0.4$, $\Delta$AIC$=0.9$, and $\Delta$AIC$=1.6$ which indicates a strong support for the cuscuton-like model.

\begin{table*}
\centering
\caption{$68\%$ confidence limits for the background parameters. For comparison purposes, on the superior block, we show the constraints for the $\Lambda$CDM model, while on the second one, we present the constraints for the cuscuton-like model. The results were obtained using different combinations of data sets, i.e. Cosmic Chronometers (HD), Type IA Supernova from Pantheon+ compilation (SN), BAO, and geometric CMB. The $*$ stands for a derived parameter.}
{\begin{tabular}{c c c c c c c}
\hline
\hline
 {model/dataset} & {$\Omega_b h^2$}  & {$\Omega_m$} & {$H_0$[km s$^{-1}$Mpc$^{-1}$]}  & $\omega*$ & $B$ & $2\sqrt{8}\alpha*$
\\
\hline
{\bf {$\Lambda$CDM}} & & & & Fixed to $-1$ & & \\
HD & $0.0220 \pm 0.0005$ & $0.3432 \pm 0.0633$ & $66.95 \pm 0.41$ & $-$ & $-$ & $-$\\
SN+BAO & $0.0221 \pm 0.0005$ & $0.3185 \pm 0.0133$ & $68.35 \pm 0.86$ & $-$ & $-$ & $-$ \\
HD+SN+BAO & $0.0220 \pm 0.0005$ & $0.3180 \pm 0.0117$ & $68.27 \pm 0.80$ & $-$ & $-$ & $-$ \\
HD+SN+BAO+CMB & $0.0224 \pm 0.0001$ & $0.3061 \pm 0.0072$ & $67.87 \pm 0.54$ & $-$ & $-$ & $-$ \\
\hline 
{\bf Cuscuton-like} & & & & & & \\
HD & $0.0220 \pm 0.0005$ & $0.294 \pm 0.081$ & $65.5 \pm 4.1$ & $-0.79 \pm 0.09$ & $0.782 \pm 0.199$ & $55.02 \pm 5.92$ \\
SN+BAO & $0.0220 \pm 0.0005$ & $0.297 \pm 0.019$ & $65.5\pm 2.2$ & $-0.92 \pm 0.06$  & $0.488 \pm 0.192$ & $54.81 \pm 1.51$ \\
HD+SN+BAO & $0.0220 \pm 0.0005$ & $0.306 \pm 0.016$ & $66.6 \pm 1.6$ & $-0.95 \pm 0.05$ & $0.403 \pm 0.186$ & $55.45 \pm 1.17$\\
HD+SN+BAO+CMB & $0.0230 \pm 0.0005$ & $0.310 \pm 0.007$ & $67.2 \pm 0.7$ & $-0.96 \pm 0.03$ & $0.337 \pm 0.115$ & $55.81 \pm 0.89$ \\
\hline
\hline
\end{tabular} \label{tab:Tabel_results_1}}
\end{table*} 

\section{Final remarks}\label{final}

We have explored theoretical aspects of some quintessence models in the presence of a cuscuton-like contribution. We expanded the discussion on dark energy models by discussing several cosmic evolution scenarios that describe the current accelerated phase of the universe compatible with the predominance of the cosmological constant. The models studied allow us to analyze several transition scenarios between an arbitrary phase and the current $\omega=-1$ regime, including as an initial phase the predominance of matter or radiation. We have observed that the presence of the cuscuton term modifies the cosmic evolution, being able to produce, in some of the models studied, a fluctuation in the equation of state of the initial phase as well as the appearance of a phantom phase that anticipates the cosmological constant phase. We have seen that, even in a model with a single scalar field and exponential potential, which in standard dynamics leads to a constant equation of state, the presence of the cuscuton term could produce a scenario in which the equation of state evolves from the decelerated to the accelerated phase of the predominance of the cosmological constant.

We have made use of a first-order framework that allowed us to study all models analytically and analyze situations in which a field with an additional cuscuton term in its dynamics interacts with another field with standard dynamics. To better understand the interaction between the fields, we have defined a relative density of interaction and observed that this density is more significant during the transition from the initial phase of evolution to the accelerated phase, exactly in the region where there is an inversion of the predominance between the fields.

In models of interacting fields, we obtained that the cuscuton term interferes differently depending on the potential of each field. When combining two exponential potentials, the parameter $\alpha$ does not control the starting value of the equation of state or produce a phantom phase, but it does control the rate at which the transition is performed. On the other hand, when we combine two hyperbolic potentials, $\alpha$ also controls the amplitude of the fluctuation in the initial value of $\omega$ as well as giving rise to a phantom phase. The combination of the two types of potential, with the cuscuton-like contribution associated with the exponential potential, produces a scenario in which the parameter $\alpha$ controls the starting value of the equation of state, thus defining the nature of the initial phase, without initial fluctuation and no phantom phase. Finally, combining the same potentials but this time associating the cuscuton term with the hyperbolic potential, the term $\alpha$ produces and controls the amplitude of the fluctuation around the initial value of $\omega$ without a phantom phase appearing anticipating the phase of a cosmological constant. We have therefore observed that the phantom phase does not occur in any of the examples when the potential has an exponential contribution.

We presented results using observational probes to put constraints on the cuscuton-like contribution for the dark energy evolution. We have combined Cosmic Chronometers, type Ia supernovae, BAO, and CMB data in different datasets and performed an MCMC analysis using the SimpleMC code, which considers only the background evolution for the parameter estimation. The results obtained when considering the extended datasets present a good agreement with the $\Lambda$CDM model, and, the cuscuton-like model is, actually, better supported by the data according to the AIC values. We stand, although, that a more rigorous analysis should be done, perhaps considering the full CMB likelihood and also the evolution of the perturbations in order to understand the effects of the cuscuton-like term on the structure formation as well as to the full cosmological parameter space.

The results of the present study, therefore, expand the possibilities of investigation of the current accelerated evolution and open several distinct ways to explore different contexts involving non-canonical scalar fields. Among possible lines of current interest, we recall, in particular, the case of phantom dark energy, quintom dark energy, and the presence of tachyonic scalar fields. We can also think of studying cosmic evolution in other contexts, such as the scalar-tensor representation of gravity, following the lines of \cite{BBB,XXX,Lobo} and references therein. It would also be of interest to investigate the custucon-like contribution within the scope of cuscuton inflation. There are other possibilities, in particular, the case that includes the cuscuton-like contribution with the quadratic scalar curvature correction and non-minimal coupling on a quintessence model with exponential potential in the Palatini formulation of gravity \cite{IA}. Another line of current interest concerns modifications in the scalar field sector, for instance, changing the kinematics to include terms with higher-order power in the derivative of the scalar fields and also, using other forms for the potential, to add distinct solutions, looking for the possibility of describing evolutions with oscillatory phases which may lead to spontaneous breaking of time translation symmetry \cite{Wil} and give rise to other interesting cosmological scenarios \cite{AAA,Neto}. We hope to report on some of these issues in the near future.

\section*{Acknowledgements}

This work is supported by funds provided by Conselho Nacional de
Desenvolvimento Científico e Tecnológico, CNPq, Grants no. 303469/2019-6 and no.  402830/2023-7, and by Paraiba State Research Foundation, FAPESQ-PB, Grant no. 0015/2019. 
S.S.C. acknowledges support from the Fondazione Cassa di Risparmio di Trento e Rovereto (CARITRO Foundation) through a Caritro Fellowship (project ``Inflation and dark sector physics in light of next-generation cosmological surveys''), Trento Institute for Fundamental Physics and Applications - TIFPA, and of Istituto Nazionale di Fisica Nucleare - INFN, Sezione di Pisa, specific initiative TASP.

\bibliographystyle{plain}

\end{document}